\documentclass[showpacs,twocolumn,pra]{revtex4}
\usepackage{graphicx,amsfonts,amsmath,psfrag,rotating}
\begin{document}  

\title{Photon Statistics of a Single Atom Intracavity EIT-Kerr System}

\author{S. Rebi\'{c}}
\email[E-mail: ]{s.rebic@auckland.ac.nz} 
\author{A. S. Parkins}
\author{S. M. Tan}
\affiliation{Department of Physics, University of Auckland, Private Bag 92019, Auckland, New Zealand}

\begin{abstract}
We explore the photon statistics of light emitted from a system comprising a single four--level atom strongly coupled to a high-finesse optical cavity mode which is driven by a coherent laser field. In the weak driving regime this system is found to exhibit a photon blockade effect. For intermediate driving strengths we find a sudden change in the photon statistics of the light emitted from the cavity. Photon antibunching switches to photon bunching over a very narrow range of intracavity photon number. It is proven that this sudden change in photon statistics occurs due to the existence of robust quantum interference of transitions between the dressed states of the atom-cavity system. Furthermore, it is shown that the strong photon bunching is a nonclassical effect for certain values of driving field strength, violating classical inequalities for field correlations.
\end{abstract}

\pacs{42.50.-p, 32.80.-t, 42.65.-k}

\maketitle

\section{Introduction}
\label{sec:intro}

A promising avenue of research in the attempt to build quantum computers uses large optical nonlinearities to create essential elements for quantum computation, such as quantum gates~\cite{Nielsen00}. Obtaining large, noiseless nonlinearities is of course a fundamental goal in the field of nonlinear optics, and technological advances in recent years, for example in cavity quantum electrodynamics (CQED)~\cite{Berman94} and the ability to access and study strongly coupled quantum systems, offer an exciting new era in this field.

Attempts to achieve large optical nonlinearities are generally plagued by a seemingly necessary trade-off between large absorption and the strength of nonlinearity: to obtain a strong nonlinearity one has to drive the atomic system close to resonance, triggering large absorption and therefore a strong source of noise. However, it is possible to reduce atomic absorption (and hence spontaneous emission) by utilizing quantum coherence effects in multilevel atoms. In particular, electromagnetically induced transparency (EIT)~\cite{Harris97} was employed by Schmidt and Imamo\u{g}lu~\cite{Schmidt96} to devise a scheme involving four--level atoms which produces a large Kerr nonlinearity with virtually no noise. It was then shown by Imamo\u{g}lu \textit{et al.}~\cite{Imam97} that if such a strong optical Kerr nonlinearity is implemented in a CQED setting, then it is possible to realize the effect of {\it photon blockade}, in which the atom--cavity system effectively acts as a turnstile device for single photons. The physical explanation of this effect is simple. Only the transition between the ground and first excited state of the nonlinear atom--cavity system is resonant with the driving field. Higher states are detuned from resonance by an amount proportional to the strength of nonlinearity. Tian and Carmichael~\cite{Tian92} have also predicted such an effect with a single two-level atom strongly coupled to the cavity mode. 

The proposal of Schmidt and Imamo\u{g}lu~\cite{Schmidt96}, although very appealing in its use of EIT to reduce decoherence, meets obstacles in the attempt for implementation in the many-atom CQED environment. It was shown that this particular realization of the photon blockade system is not ideal for demonstrating photon blockade, because of the strong linear dispersion of the medium~\cite{Grangier98,Gheri99}. In an attempt to avoid this difficulty, Rebi\'{c} \textit{et al.}~\cite{Rebic99} proposed a model in which a single four--level atom is trapped in a high--finesse microcavity. It was shown that this system (called the single-atom EIT-Kerr system) can act as a near-ideal Kerr optical nonlinearity. In such a strongly coupled system, the excitations can be labeled as polaritons, which are defined as mixtures of atom/cavity mode excitations. For weak to moderate driving, the EIT--Kerr system can be approximated by a two-state system, corresponding to the two lowest lying polariton eigenstates. The transitions to the next set of excited states (the second manifold), obtained by introducing a second photon from the driving field into the cavity, are highly detuned from the bare--cavity resonance, and therefore cannot be excited. Hence, in effect, the weakly driven single--atom EIT--Kerr system acts as an ideal photon blockade device.

In a further work~\cite{Rebic01}, a Hamiltonian for the effective two-level system was derived using a polariton basis, and it was shown that such a Hamiltonian can describe the spectral properties and energy splittings (dynamic Stark effect) seen in the full model. Furthermore, to explain the properties of the second order correlation function, it was shown that the effective two-level system has to be generalized to include two extra states in the first excitation manifold~\cite{Rebic00}. If more than one atom is introduced the situation becomes more complicated, since each additional atom introduces extra energy levels into the second manifold, some of which are connected to the lower levels via transitions which are resonant with the cavity mode. However, it was shown by Werner and Imamo\u{g}lu~\cite{Werner99} that the introduction of an additional atomic detuning can be used to solve this problem (see also the work of Greentree \textit{et al.}~\cite{Green00}).

In~\cite{Rebic00}, a comparison of the EIT--Kerr scheme and the extended Jaynes--Cummings scheme (i.e. the standard Jaynes-Cummings model~\cite{JCmodel} with the effects of losses included) was made in terms of their effectiveness in producing photon blockade. The EIT--Kerr scheme exhibited a value of $g^{(2)}(0)$ (the second order correlation function at zero time delay) of $\frac{1}{40}$ that of the corresponding Jaynes--Cummings scheme. This was attributed to destructive quantum interference between certain transition amplitudes. In particular, with a suitable choice of parameters, probability amplitudes for transitions from the first excited state (in the first manifold) to the two second manifold eigenstates closest to the cavity resonance cancel each other. This leads to enhanced antibunching of the photons leaving the cavity so that $g^{(2)}(0) \approx 0$.

We have remarked that the validity of these results depends on the assumption of weak driving. It is therefore of interest to explore how robust the photon blockade is as the driving field strength is increased. We have performed an analysis of the driving field dependence in the EIT--Kerr system, and briefly compared the results with those obtained for the extended Jaynes--Cummings model. A surprising feature is found in the EIT--Kerr system, where a sudden change in photon statistics (i.e. of $g^{(2)}(0)$) of light emerging from the cavity occurs at intermediate driving strengths. If plotted against the mean intracavity photon number, $g^{(2)}(0)$ is seen to undergo a sudden transition, switching rapidly between antibunching and strong bunching. The exact position of the threshold depends on the characteristic system parameters, namely the atom--field interaction strength and the (EIT) coupling field Rabi frequency. 

In Section II we present our model and the methods of solution employed. Section III contains the results of numerical simulations for the photon statistics, showing the surprising behaviour of the second order correlations in the regime of intermediate driving. Section IV contains a detailed explanation for the results presented in the preceding Section, obtained by analyzing the density matrix expressed in an appropriate basis. In Section V we present analytical calculations based on insights obtained in Section IV.

\section{Model}
\label{sec:model}

\subsection{Hamiltonian}

\begin{figure}[!t]
   %\vspace{1cm}
   \includegraphics[scale=0.6]{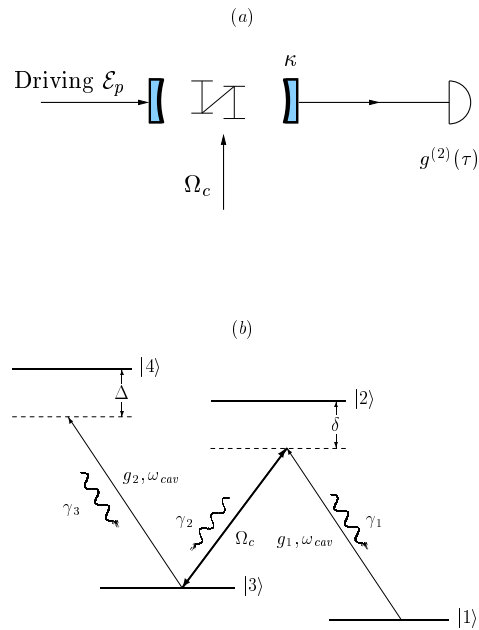}
  %\vspace{2cm}
   \caption{$(a)$ Envisaged experimental setup. A single four-level atom is trapped in an optical cavity with the decay constant $\kappa$. The cavity is driven by a coherent laser field which couples to the cavity mode with strength ${\mathcal E}_p$. An additional laser with Rabi frequency $\Omega_c$ directly couples to the atomic transition. $(b)$ Atomic energy level scheme. The cavity mode drives transitions $|1\rangle \rightarrow |2\rangle$ and $|3\rangle \rightarrow |4\rangle$, with the respective coupling strengths $g_1$ and $g_2$. The transition $|2\rangle \leftrightarrow |3\rangle$ is coupled by a classical field of frequency $\omega_c$ and Rabi frequency $\Omega_c$. Spontaneous emission rates are denoted by $\gamma_j$. Detunings $\delta$ and $\Delta$ are positive in the shown configuration.}
   \label{fig:atom}
\end{figure}

The driven atom-cavity configuration and the atomic energy level scheme are shown schematically in Fig.~\ref{fig:atom}. The Hamiltonian describing the system in the rotating wave and electric dipole approximations is $\mathcal{H} = \mathcal{H}_0 + \mathcal{H}_d$, where, in the interaction picture,
\begin{subequations}
 \label{eq:hamiltonian}
 \begin{eqnarray}
  \mathcal{H}_0 &=& \hbar\delta \, \sigma_{22} + \hbar\Delta \, \sigma_{44} + i\hbar g_1\, \bigl( a^\dagger \sigma_{12} - \sigma_{21} a \bigr) \nonumber \\ 
  &\ & + i\hbar \Omega_c\, \bigl( \sigma_{23} - \sigma_{32} \bigr) + i\hbar g_2\, \bigl( a^\dagger \sigma_{34} - \sigma_{43} a \bigr) , \label{eq:h0} \\
  \mathcal{H}_d &=& i\hbar \mathcal{E}_p \, \bigl( a - a^\dagger \bigr). \label{eq:hd} 
 \end{eqnarray}
\end{subequations}
In these equations, $\sigma_{ij}$ are atomic raising and lowering operators (for $i \neq j$), and atomic energy level population operators (for $i = j$); $a$ and $a^\dagger$ are the cavity field annihilation and creation operators, respectively. Detunings $\delta$ and $\Delta$ are defined from the relevant atomic energy levels; $g_{1,2}$ are atom--field coupling constants for the transitions $|1\rangle \rightarrow |2\rangle$ and $|3\rangle \rightarrow |4\rangle$, and $\Omega_c$ is the Rabi frequency of a coherent field coupling the transition $|2\rangle \rightarrow |3\rangle$. The cavity driving field is characterized through the parameter ${\mathcal E}_p$, related to the power output of the driving laser $\mathcal{P}$ via
\begin{equation}
 \label{eq:ep}
 \mathcal{E}_p = \sqrt{\frac{\mathcal{P} \kappa T^2}{4\hbar\omega_{cav}}},
\end{equation}
where $T$ is the cavity mirror transmission coefficient, $\kappa$ is the cavity decay rate, and $\omega_{cav}$ is the cavity mode frequency. Damping due to cavity decay and spontaneous emission is introduced below in the context of the quantum trajectory approach~\cite{Carmichael93}.

Four atomic levels plus the cavity mode span a Hilbert space of dimension $4 \times N$, where $N$ denotes the truncation order in a Fock state expansion of the cavity field subspace. In the absence of driving, Hamiltonian ${\mathcal H}_0$, given by Eq.~(\ref{eq:h0}), takes a block--diagonal form, with $N$ blocks on the main diagonal. Each block represents a manifold of eigenstates associated with the appropriate term in the Fock expansion. The ground, first and second manifolds have been analyzed from the viewpoint of photon blockade in Refs~\cite{Rebic99,Werner99,Green00}. The general structure of the dressed states in an arbitrary $n^{\rm{th}}$ manifold has been discussed in Ref.~\cite{Rebic01}.

Addition of the driving term~(\ref{eq:hd}) complicates the analysis, since the Hamiltonian matrix does not retain the block--diagonal form. It is possible, however, to re--express the driving Hamiltonian in terms of polariton operators and effective Rabi frequencies for transitions between dressed states. This has been done in~\cite{Rebic01}, and in the remainder of this article we will draw on these results.

Our analysis of the driven `atom-cavity molecule' will proceed in two complementary directions. First, using the method of quantum trajectories~\cite{Carmichael93}, we obtain numerical data. Then, using the formalism of Ref.~\cite{Rebic01}, we construct an effective Hamiltonian in the polariton basis, which encapsulates the physics sufficiently to explain the numerical results.

\subsection{Damping: Quantum Trajectories}

In the quantum trajectories approach, damping enters the model through collapse operators, with each of these corresponding to one decay channel. In the case under consideration we require the following four collapse operators,
\begin{eqnarray}
 C_1 &=& \sqrt{\gamma_1}\, \sigma_{12}, \ \ C_2 = \sqrt{\gamma_2}\, \sigma_{32}, \nonumber \\
 C_3 &=& \sqrt{\gamma_3}\, \sigma_{34}, \ \ C_4 = \sqrt{\kappa}\, a\, ,
\end{eqnarray}
where $\gamma_k$ denote spontaneous emission rates into each of the decay channels, and $\kappa$ is the cavity field decay rate. The effective Hamiltonian used in the trajectories approach is non--Hermitian and takes the form
\begin{equation}
 {\mathcal H}_{eff} = {\mathcal H} - i \sum_{k=1}^4 C_k^\dagger C_k \, , \label{eq:heff}
\end{equation}
with ${\mathcal H}$ given by~(\ref{eq:hamiltonian}).

In deciding on the truncation for the cavity mode Hilbert space, special care must be taken to include a sufficient number of states to capture all the relevant dynamics. If an empty cavity would be driven by an external coherent field corresponding to the parameter ${\mathcal E}_p$, the amplitude of the intracavity coherent field would be
\begin{equation}
 \alpha = i{\mathcal E}_p/\kappa \, ,
\end{equation}
and the expected mean intracavity photon number $\langle n \rangle = |\alpha|^2$. Our simulations include driving strengths of up to ${\mathcal E}_p = 3\kappa$, so the truncation is set at $N = 40$. The inset in Fig.~\ref{fig:1} shows that the actual mean intracavity photon number stays well below its empty cavity counterpart for the given range of driving, thus justifying the chosen truncation. 

\section{Numerical Simulations of Photon Statistics}
\label{sec:numsim}

In this section we present the results of numerical simulations using the method of quantum trajectories~\cite{Carmichael93}, with results averaged over 10000 trajectories. Values of parameters chosen for the simulations are consistent with recent experiments~\cite{Hood00}, so the scheme presented in this paper should be experimentally viable.

We evaluate the second--order correlation function $g^{(2)}(0)$ as a function of driving strength ${\mathcal E}_p$. It was established earlier~\cite{Imam97,Rebic99} that this function is a good measure of photon blockade; perfect photon blockade yields perfectly antibunched photons. The steady--state second--order correlation function is given by
\begin{equation}
\label{eq:g2}
 g^{(2)}(\tau) = \lim_{t \rightarrow \infty} \frac{\langle a^\dagger (t) a^\dagger (t+\tau) a(t+\tau) a(t)\rangle}{\langle a^\dagger (t) a(t)\rangle \langle a^\dagger (t+\tau) a(t+\tau) \rangle} \, .
\end{equation}
Perfect antibunching/photon blockade corresponds to $g^{(2)}(\tau=0) = 0$; for a coherent field $g^{(2)}(\tau=0) = 1$, and for a bunched field $g^{(2)}(\tau=0) > 1$~\cite{Walls94}.

\begin{figure}[!t]
   \vspace{1cm}
   \includegraphics[scale=0.52]{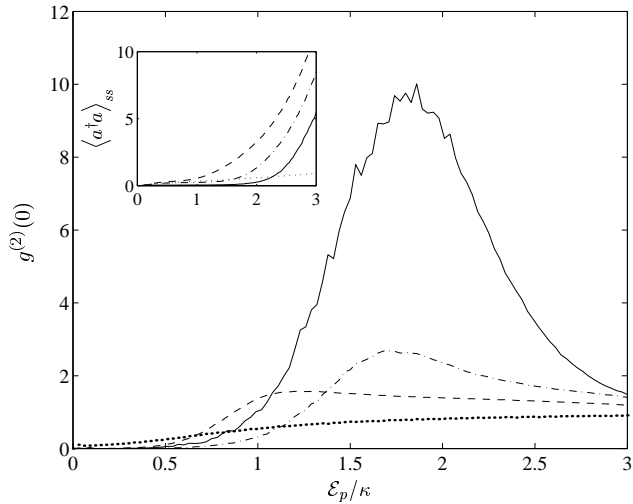}
  %\vspace{2cm}
   \caption{Second order correlations at zero time--delay against the (scaled) driving parameter. The inset shows mean intracavity photon number in the steady state. The dotted line denotes the extended Jaynes--Cummings scheme for which $\gamma = 0.1\kappa$, $g = 6 \kappa$. Other curves show the single--atom EIT--Kerr model with $\gamma_j = 0.1\kappa$, $g_l = 6 \kappa$, $\delta = -0.2\kappa$, $\Delta = 0.1\kappa$. The solid line represents the case $\Omega_c = 2\kappa$; dot--dashed line $\Omega_c = 6\kappa$; dashed line $\Omega_c = 18\kappa$. All curves are obtained by averaging over $10^4$ trajectories.}
   \label{fig:1}
\end{figure}

Fig.~\ref{fig:1} shows $g^{(2)}(\tau=0)$ as a function of the (scaled) driving parameter for a single--atom EIT--Kerr system, compared with the same quantity calculated for the extended Jaynes--Cummings model. The extended JC model is driven on the lower Rabi resonance, as envisaged by Tian and Carmichael~\cite{Tian92}. In the weak driving regime, simulation confirms earlier results -- stronger photon blockade in the EIT--Kerr system. As the driving increases, the extended Jaynes-Cummings model gradually saturates, and the field correlation asymptotically tends to one. Naively, one would expect qualitatively similar behavior in the EIT--Kerr model. Our simulation, however, shows a vastly different result. Over a narrow range of driving, the statistics of the field changes from strongly antibunched to strongly bunched, and the coherent state value $g^{(2)}(0) = 1$ is approached asymptotically from above as ${\mathcal E}_p$ is increased further. 

\begin{figure}[!t]
   \includegraphics[scale=0.75]{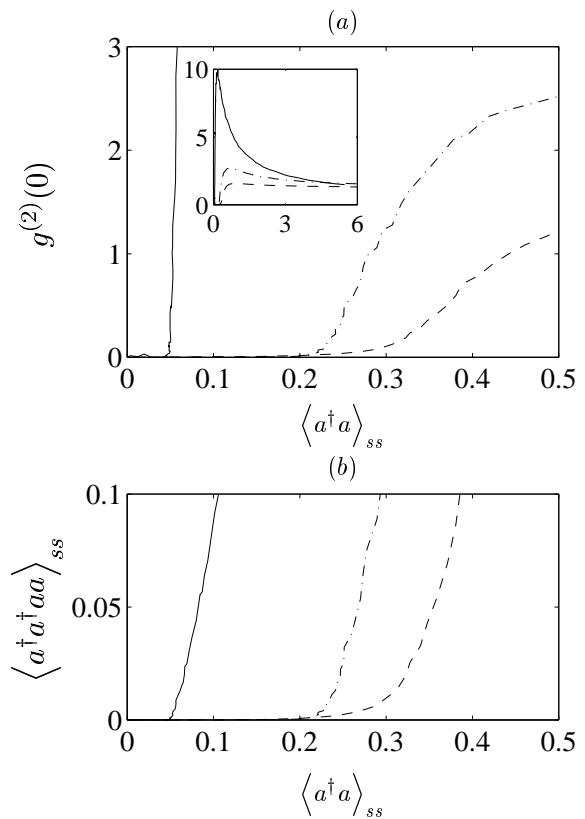}
   %\vspace{2cm}
   \caption{Second order correlations at zero time--delay against the mean steady--state intracavity photon number for a single--atom EIT--Kerr system. Parameters are the same as in Fig.~\ref{fig:1}. Figure $(a)$ shows normalized correlation function with main figure showing the range where the sudden change can be seen in detail, while the inset shows the whole range. Figure $(b)$ shows the unnormalized second order correlations.}
   \label{fig:2}
\end{figure}

Note that the quantity being increased here is the \textit{scaled} driving parameter, so one may argue that it is the ratio that determines this behavior, i.e. we may either increase the driving strength or decrease the cavity decay rate to obtain the same result. However, to clarify this issue as related to the photon statistics, in Fig.~\ref{fig:2} we show a parametric plot of second--order correlation function against the expectation value of intracavity photon number, $\langle a^\dagger a\rangle_{ss}$. Both quantities are now unscaled by any external parameter, and the sudden nature of the switch becomes even more obvious, showing a phase--transition--like behaviour. This can be related back to Fig.~\ref{fig:1}. In particular, the transition from antibunching to bunching happens over a range of driving for which the intracavity photon number stays practically constant (see inset of Fig.~\ref{fig:1}), leading to the suddenness of the transition seen in Fig.~\ref{fig:2}, and the concentration of points around the threshold region. To further emphasize this, in Fig.~\ref{fig:2} $(b)$ we plot the `pure' second--order correlation function $\langle a^\dagger a^\dagger a a\rangle_{ss}$ which also exhibits a threshold-like behaviour.

One other feature of the numerical results is noted. The bunching--antibunching transition is sharper and the subsequent bunching stronger for smaller $\Omega_c$. In fact, it is the increasing ratio $g_1/\Omega_c$ that really matters. The sharper transitions also occur at smaller values of $\langle a^\dagger a\rangle_{ss}$. For a decreasing ratio $g_1/\Omega_c$, $g^{(2)}(0)$ approaches the behaviour of the extended JC model.

\section{Density Matrix Treatment}
\label{sec:densmat}

We proceed to determine which eigenstates of the strongly coupled quantum system contribute significantly to its dynamics. The total density matrix of the system in the steady state can be written in most general terms as
\begin{equation}
  \label{eq:gendensity}
  \rho = \sum_{a,b} c_{ab}|a\rangle\langle b|\, ,
\end{equation}
where $a$, $b$ belong to the set of all possible states of the system in an arbitrary basis. The natural basis for the simulation is the one of bare states. The cavity mode subspace is truncated at 40, making the size of $\rho$ equal to $160\times 160$. Let $T$ be the transformation that diagonalizes Hamiltonian ~(\ref{eq:hamiltonian}), i.e. the full Hamiltonian, with driving included. The density matrix can be transformed into a new basis as $\rho_{diag} = T\rho T^{-1}$, and we can look for the nonzero elements of this matrix. Diagonal elements of the matrix correspond to populations of the dressed states, while off-diagonal elements correspond to coherences between the dressed states. The results depend on the size of $\mathcal{E}_p$, i.e. we expect the number of non--zero elements to increase as $\mathcal{E}_p$ is increased. For a large part of the $\mathcal{E}_p$ range considered, however, the density matrix is dominated by elements corresponding to two particular states.

\begin{figure*}[!t]
  \begin{center}
  \includegraphics[scale=0.9]{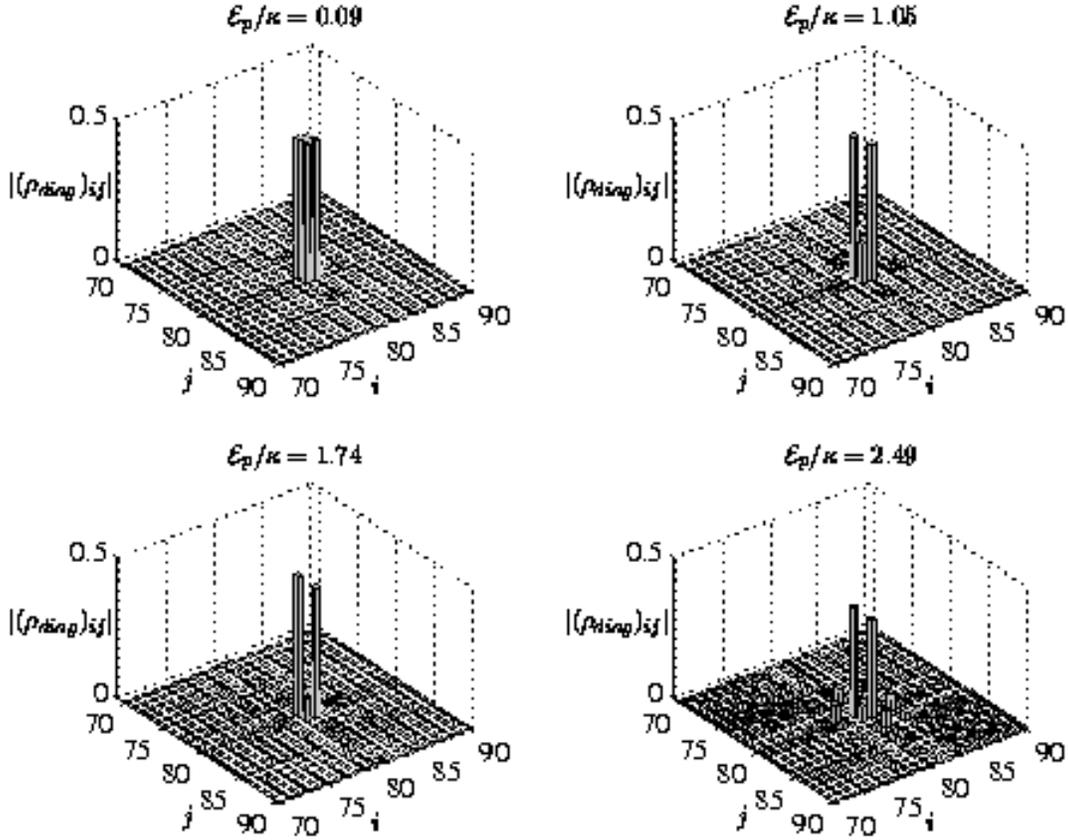}
  \end{center}
  \caption{Relevant density matrix elements at the different values of driving. Elements of the central submatrix (dimension $21 \times 21$) of the total density matrix ($160 \times 160$) are shown. In the basis which diagonalises the Hamiltonian, the chosen submatrix contains elements corresponding to the dressed states closest to the cavity resonance.}
  \label{fig:denmat}
\end{figure*}

Figure~\ref{fig:denmat} shows the nonzero matrix elements of density matrix $\rho_{diag}$. Diagonal states are sorted in increasing order; the state with largest negative detuning is at $(i,\, j) = (1,\, 1)$, while the state with largest positive detuning is at $(i,\, j) = (160,\, 160)$. States with the smallest detuning (i.e. closest to the cavity resonance) are at the centre of matrix, at positions 80 and 81 along the main diagonal. Two diagonal elements dominate the matrix, and we identify these as being the populations of the Stark-split states $|\psi_\pm \rangle$ (see Appendix~\ref{sec:stark} and~\cite{Rebic01}). Stark-split states are therefore found at the positions 80 and 81 on the main diagonal of the density matrix. At positions 79 and 82 are the two states from the second manifold (two-photon excitations), closest to the resonance. At the positions 78 and 83 are two remaining states from the first manifold; at the positions 77 and 84 are two states from the third manifold (three-photon excitations), closest to the resonance, and so on. The states we have just identified suffice to indicate the dynamics of the system. Off-diagonal elements are coherences between the appropriate dressed states.

One striking feature can be noted immediately from Fig.~\ref{fig:denmat}. Namely, the first `square' of elements (4 elements on each side) encircling the centre square (Stark states, $2\times 2$) remains much smaller than all of the other accessible states. In terms of dressed states, this means that the second manifold states remain unpopulated and the coherences between these and all the other states vanish as well. Furthermore, as the driving increases, the states whose coherences with the Stark states increase to a noticeable size are the third manifold states (see Fig.~\ref{fig:denmat} for ${\mathcal E}_p = 0.09\kappa$). As the driving increases further, the remaining two first manifold states (their populations and coherences with the Stark states) also start to contribute. So, contrary to expectations, the contribution to the dynamics of the states beyond the effective two-level system do not increase according to the smallness of their detunings from the bare cavity frequency. The two least--detuned second-manifold states are in fact essentially unpopulated.

From the above analysis of the density matrix elements, we can deduce the solution to the `photon statistics puzzle' of Figs.~\ref{fig:1} and~\ref{fig:2}. The strong photon antibunching at small ${\mathcal E}_p$ is the consequence of the effective two-level system, and has been well understood. The sharp rise in $g^{(2)}(0)$ with increasing ${\mathcal E}_p$ can be attributed to the two-photon transitions needed to populate the third manifold states from the first manifold, and then decay back in cascade to the first manifold. These two-photon decays cause the sharp increase in the second order output field correlations. But, what causes the system to skip second manifold states? Again, the answer can be deduced from the density matrix. Strong coherence between the first and third manifold states, together with the missing population in the second manifold states (and vanishing coherences associated with these states) uncovers the effect of EIT-type quantum interference between the dressed states. This is not surprising, since the quantum interference between the transitions from first to second manifold have been already discussed in Ref.~\cite{Rebic00}.

\begin{figure}[!t]
  \begin{center}
  \includegraphics[scale=0.65]{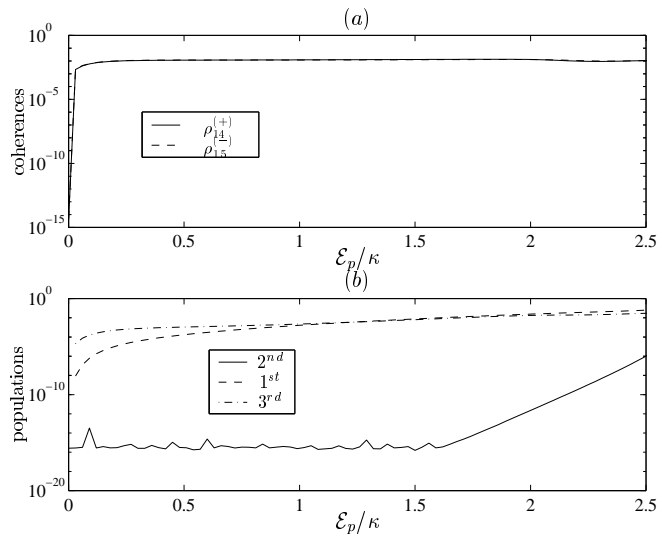}
  \end{center}
  \caption{Semilogarithmic plots of the density matrix elements. $(a)$ Plot of the coherences between the first manifold and third manifold states. Coherence $\rho_{14}^{(+)}$ is the coherence between the upper Stark state and lower third manifold state. Coherence $\rho_{15}^{(-)}$ is the coherence between the lower Stark state and upper third manifold state. $(b)$ Plot of the populations of the dressed states. $2^{nd}$ denotes a sum of populations in the two relevant states of the second manifold, $1^{st}$ denotes a sum of populations in the two far detuned states in the first manifold and $3^{rd}$ denotes a sum of populations in the two relevant states of the third manifold. Parameters are as in Fig.~\ref{fig:1}, solid line.}
  \label{fig:popcoh}
\end{figure}

These features are shown in Fig.~\ref{fig:popcoh}, where the relevant populations and coherences are shown. Note that the combined population of the two inner second manifold states (Fig.~\ref{fig:popcoh} $(b)$, solid line) vanishes for a large interval of driving, since the value of $\sim 10^{-16}$ is at the numerical precision value, and fluctuations are numerical, not physical in nature. These populations become nonzero at the value of driving strength at which $g^{(2)}(0)$ of Fig.~\ref{fig:1} (solid line) peaks. These plots further justify the discussion of the preceeding paragraphs.

We proceed with the development of an effective model with relatively few levels which nevertheless captures most of the important features of the dynamics.

\section{Effective Model}
\label{sec:qint}

\begin{figure}[!t]
  \begin{center}
  \includegraphics[scale=0.8]{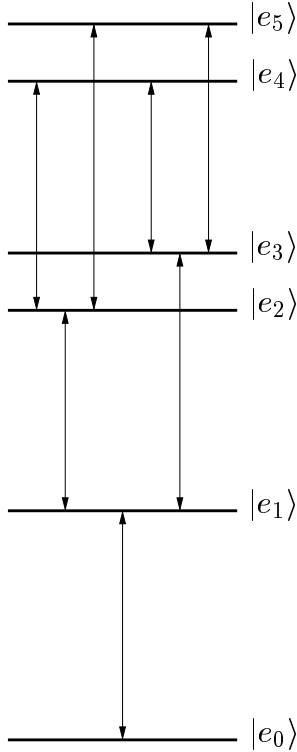}
  \end{center}
  \caption{Schematic depiction of the six states used to formulate the effective model. States $|e_0\rangle$ and $|e_1\rangle$ are the effective two-level system from Ref.~\cite{Rebic01}. States $|e_2\rangle$ and $|e_3\rangle$ are second manifold states closest to the bare cavity resonance, and $|e_4\rangle$ and $|e_5\rangle$ are the third manifold states closest to the resonance. Arrows represent effective driving.}
  \label{fig:eff6}
\end{figure}

In the formulation of an effective model, we rely on the formalism developed in Ref.~\cite{Rebic01}. This formalism was very successful in the development of an effective two-level theory, and in explaining the fluorescence spectrum. Now we extend the model and include a total of six dressed states in the effective model. These states are shown in Fig.~\ref{fig:eff6}.

It should be noted, however, that the dressed states shown in Fig.~\ref{fig:eff6} do not correspond exactly to the dressed states discussed in Section~\ref{sec:densmat}. Namely, the dressed states of Sec.~\ref{sec:densmat} are often referred to as doubly dressed states, since they diagonalize the Hamiltonian with driving contributions included. In this Section, we will treat driving separately (see~\cite{Rebic01}), and let the dressed states represent the eigenstates of the interaction Hamiltonian~(\ref{eq:h0}) alone. Driving can then be included through effective Rabi frequencies $\Omega_{ij}$, coupling dressed states $|e_i\rangle$ and $|e_j\rangle$.

The relevant effective non-Hermitian Hamiltonian (including driving and damping) is thus
\begin{eqnarray}
  \label{eq:heff6}
  {\mathcal H}_{eff} &=& \hbar\epsilon_2 \, p_2^\dagger p_2 + \hbar\epsilon_3 \, p_3^\dagger p_3 + \hbar\epsilon_4 \, p_{j4}^\dagger p_{j4} + \hbar\epsilon_5 \, p_{j5}^\dagger p_{j5} \nonumber \\
  &\ &+i\hbar \Omega_{01} \left( p_1 - p_1^\dagger \right) \nonumber \\
  &\ &+ i\hbar \left( \Omega_{12}^* p_2 - \Omega_{12} p_2^\dagger \right) + i\hbar \left( \Omega_{13}^* p_3 - \Omega_{13} p_3^\dagger \right) \nonumber \\
  &\ & + i\hbar \Omega_{24}  \left( p_{24} -  p_{24}^\dagger \right) + i\hbar \Omega_{25} \left( p_{25} -  p_{25}^\dagger \right) \nonumber \\
  &\ & + i\hbar \Omega_{34}  \left( p_{34} -  p_{34}^\dagger \right) + i\hbar \Omega_{35} \left( p_{35} -  p_{35}^\dagger \right) \nonumber \\
  &\ & - i\hbar \Gamma_0 p_1^\dagger p_1 - i\hbar \Gamma_{22} p_2^\dagger p_2 - i\hbar \Gamma_{33} p_3^\dagger p_3 \nonumber \\
  &\ & - i\hbar \Gamma_{44} p_{j4}^\dagger p_{j4} - i\hbar \Gamma_{55} p_{j5}^\dagger p_{j5} \nonumber \\
  &\ &- i\hbar \Gamma_{23} p_{2}^\dagger p_{3} - i\hbar \Gamma_{32} p_{3}^\dagger p_{2} \nonumber \\
  &\ &- i\hbar \Gamma_{45} p_{j4}^\dagger p_{j5} - i\hbar \Gamma_{54} p_{j5}^\dagger p_{j4} \, .
\end{eqnarray}
Operators $p_{jk}$ are polariton operators defined by $|e_k\rangle = p_{jk}^\dagger |e_j\rangle$. For the states $|e_k\rangle$ which are accessible from only one lower state, the notation has been abbreviated, so, for example, $p_2 \equiv p_{12}$. Note also that the indices $j$ are dummy indices, i.e. $p_{jk}^\dagger p_{jl} = |e_k\rangle\langle e_l|$. Rabi frequencies $\Omega_{12}$ and $\Omega_{13}$ have a phase term making them purely imaginary, while all the other Rabi frequencies are real. Damping terms were discussed in detail in~\cite{Rebic01}. It is now possible to formalize the distinction between the two dressed states bases used in this Section and Section~\ref{sec:densmat}. The basis in which the Hamiltonian~(\ref{eq:heff6}) is written is defined in Ref.~\cite{Rebic01} and depicted in Fig.~\ref{fig:eff6}. The discussion of Section~\ref{sec:densmat} is based upon diagonalizing the total Hamiltonian (including all of the 160 states used for the numerical simulation).

With the effective Hamiltonian~(\ref{eq:heff6}), we can also write the master equation for an effective density matrix as
\begin{eqnarray}
  \dot{\rho}_{eff} &=& -\frac{i}{\hbar} \left( {\mathcal H}_{eff}\rho_{eff} - \rho_{eff}{\mathcal H}_{eff}^\dagger \right) \nonumber \\
  &\ &+2\sum_{i,j} S_i\rho_{eff} S_j^\dagger   \label{eq:master} \, ,
\end{eqnarray}
where $S_k$ now denote the polariton collapse operators (Appendix~\ref{sec:jump}). The effective density matrix $\rho_{eff}$ has the dimension $6 \times 6$ - a significant reduction from $160 \times 160$ used to obtain the results in Sec.~\ref{sec:densmat}. Equations of motion for the elements of $\rho_{eff}$ are given in Appendix~\ref{sec:rhoij}.

From the equations of motion, we can uncover terms which lead to the effect of quantum interference. For example, equations for the populations of the second manifold states $\rho_{22}$ and $\rho_{33}$, depend on the populations $\rho_{44}$ and $\rho_{55}$ of the third manifold states and the {\it coherences} between these states $\rho_{45}$ and $\rho_{54}$; the latter with rate $\Gamma_{45}$. At the same time, equations for $\rho_{44}$ and $\rho_{55}$ do not depend on the second manifold states, nor their mutual coherence. The same holds for the coherences $\rho_{23}$ and $\rho_{45}$ and their adjoints; the equation for $\rho_{23}$ depends on $\rho_{44}$, $\rho_{55}$, $\rho_{45}$ and $\rho_{54}$, but not vice versa. Population of and coherence between the second manifold states is linked to the population of and coherence between the third manifold states. If this dependence is removed from the equations of motion, i.e., terms dependent on $\rho_{44}$, $\rho_{55}$, $\rho_{45}$ and  $\rho_{54}$ are removed from the equations for $\dot{\rho}_{22}$, $\dot{\rho}_{33}$, $\dot{\rho}_{23}$ and $\dot{\rho}_{32}$, cancellation of the population in the second manifold ceases to occur. 

The 35 equations of the effective model can, in principle, be solved analytically in the steady state. However, the resulting expressions are complicated and do not offer significant physical insight, so we have opted to perform numerical solutions of the equations given in Appendix~\ref{sec:rhoij} and check for validity of the effective model. Having the solutions for the populations and coherences, the second order correlation function for zero time delay can be calculated as the ratio of 
\begin{subequations}
  \label{eq:secan}
\begin{eqnarray}
  &\, &\langle a^\dagger a^\dagger a a \rangle = |w_{01}|^2 |w_{12}|^2 \rho_{22} + |w_{01}|^2 |w_{13}|^2 \rho_{33} \nonumber \\
  &\, &+\left[ |w_{12}|^2 |w_{24}|^2 + |w_{13}|^2 |w_{34}|^2 +\left(w_{12}^*w_{24}^*w_{13}w_{34}+ {\rm c.c.}\right) \right] \rho_{44} \nonumber \\
  &\, &+\left[ |w_{12}|^2 |w_{25}|^2 + |w_{13}|^2 |w_{35}|^2 +\left( w_{13}^* w_{35}^* w_{12} w_{25}+ {\rm c.c.} \right) \right] \rho_{55} \nonumber \\
  &\, &+\left( |w_{01}|^2 w_{12}w_{13} \, \rho_{34} + {\rm c.c.} \right) \nonumber \\
  &\, &+\left[ \left(|w_{12}|^2 w_{24}w_{25}^* + |w_{13}|^2 w_{34}w_{35}^* \right. \right.\nonumber \\
  &\, &\left. \left. + w_{12}^*w_{25}^*w_{13}w_{34}+ w_{12}w_{24}w_{13}^*w_{35}^* \right) \rho_{45} + {\rm c.c.}\right] \, ,
\end{eqnarray}
and the square of
\begin{eqnarray}
  &\, &\langle a^\dagger a \rangle = |w_{01}|^2\rho_{11} + |w_{12}|^2\rho_{22} + |w_{13}|^2\rho_{33} \nonumber \\
  &\, &+ \left(|w_{24}|^2 + |w_{34}|^2 \right)\rho_{44} + \left(|w_{25}|^2 + |w_{35}|^2 \right)\rho_{55} \nonumber \\
  &\, &+ \left[ w_{12} w_{13}^* \rho_{23} + \left( w_{24} w_{25}^* + w_{34} w_{35}^* \right) \rho_{45} + {\rm c.c.} \right]
\end{eqnarray}
\end{subequations}
where $w_{ij} = \Omega_{ij}/{\mathcal E}_p$, and c.c. stands for complex conjugate.

\begin{figure}
  \begin{center}
    \includegraphics[scale=0.67]{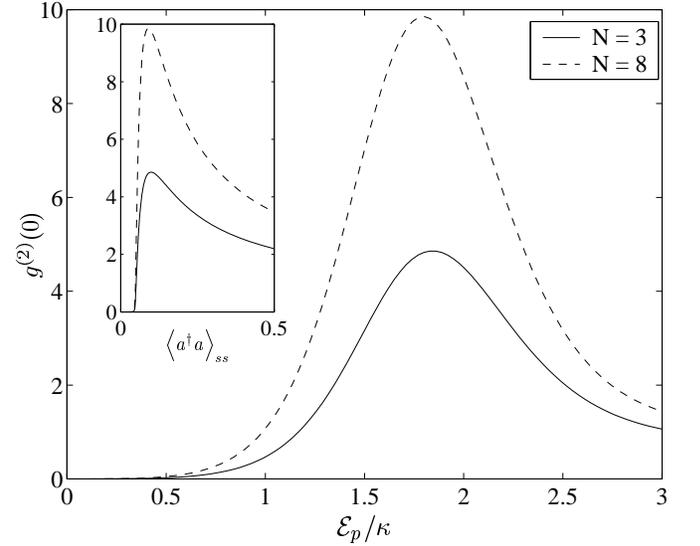}
  \end{center}
  \caption{Comparison of the numerical results obtained from the effective model including three manifolds with the results of numerical simulations including up to eight manifolds. Details are described in the text. The parameters are the same as for the solid line in Fig.~\ref{fig:1}.}
  \label{fig:comp}
\end{figure}

The results are displayed in Fig.~\ref{fig:comp}. We have compared the numerical solutions of the effective model with the results of Sec.~\ref{sec:numsim} and found a very good qualitative agreement. We do find strong bunching and threshold behaviour occurring at the same values of ${\mathcal E}_p$ and $\langle a^\dagger a \rangle_{ss}$. However, the effective model differs from the full simulations in the size of $g^{(2)}(0)$ at its peak, by approximately a factor of 2. Including more states in the effective model would yield full agreement with the numerical data. In particular, we found that including two states closest to the resonance from up to eight manifolds reproduces the numerical data exactly. The reason is that the two-photon cascade decay responsible for the behaviour of $g^{(2)}(0)$ can result from the decay of the higher states to third manifold states first, i.e. two-photon cascade can, in the manner of speaking, be driven ``from below'' and ``from above''. Naturally, the decay of higher lying states introduces more single-photon transitions as well. Therefore, adding one manifold at a time to an effective model reveals that the increase in maximum value of $g^{(2)}(0)$ gradually diminishes with new manifolds added, settling at its maximum value after the inclusion of the eighth manifold. On the other end, the effective model of Fig.~\ref{fig:eff6} is the smallest possible model which (at least qualitatively) reproduces the strong bunching effect in this system.

\section{Dynamics of the Forward Scattering of Light}
\label{sec:incoh}

To obtain a different and useful perspective on the physical processes involved in the changing nature of the statistics of light emitted by the coupled atom-cavity system, we can split the field operator into contributions from a coherent mean amplitude and from an incoherent part~\cite{Carmichael85},
\begin{subequations}
  \label{eq:incoh}
\begin{equation}
   a = \alpha + \Delta a \, , \label{eq:split}
\end{equation}
where $\alpha \equiv \langle a \rangle$ denotes the coherent amplitude component of the intracavity field, while $\Delta a$ denotes the incoherent amplitude component $\langle \Delta a \rangle = 0$, whose emergence is the result of scattering of the cavity field by the atom. Using this decomposition, one can rewrite the expression for $g^{(2)}(0)$ as
\begin{eqnarray}
  g^{(2)}(0) -1 &=& \frac{\langle a^\dagger a^\dagger aa\rangle_{ss}}{\langle a^\dagger a\rangle_{ss}^2} \nonumber \\
  &=& \frac{\langle :\left( \alpha^*\Delta a + \alpha\Delta a^\dagger\right)^2:\rangle_{ss}}{\left( |\alpha|^2 + \langle \Delta a^\dagger \Delta a \rangle_{ss}\right)^2} \nonumber \\
  &\ & + \frac{4|\alpha| {\rm Re}\left[ \langle \Delta a^{\dagger 2}\Delta a\rangle_{ss}\right]}{\left( |\alpha|^2 + \langle \Delta a^\dagger \Delta a \rangle_{ss}\right)^2} \nonumber \\
  &\ & + \frac{\langle \Delta a^{\dagger 2}\Delta a^2\rangle_{ss} - \left( \langle \Delta a^\dagger \Delta a \rangle_{ss}\right)^2}{\left( |\alpha|^2 + \langle \Delta a^\dagger \Delta a \rangle_{ss}\right)^2} \nonumber \\
  &=& S(\Delta a) + T(\Delta a) + V(\Delta a) \, ,
\end{eqnarray}
\end{subequations}
where $: :$ denotes normal ordering, and $\langle a^\dagger a\rangle_{ss} = |\alpha|^2 + \langle \Delta a^\dagger \Delta a \rangle_{ss}$. The three terms in this expansion, denoted by {\it S, T} and $V$, have been identified by Carmichael~\cite{Carmichael85} for the case of a two-level atom. The decomposition~(\ref{eq:incoh}) shows how the behaviour of $g^{(2)}(0)$ for different values of driving field can be interpreted as the effect of self-homodyning between the coherent and incoherent components of the intracavity field~\cite{Mandel82}. From this viewpoint, it is easy to identify $S(\Delta a)$ as a term describing the squeezing in the field quadrature in phase with the driving field, $V(\Delta a)$ gives the variance in the incoherent component, and $T(\Delta a)$ describes intensity-amplitude correlations in the incoherent component. Both $V(\Delta a)$ and $T(\Delta a)$ are determined by the correlations in the intensity, so the departure from the coherent value of the correlation function can be assigned to the effects of squeezing and the effects of intensity correlations.

\begin{figure}
  \begin{center}
    \includegraphics[scale=0.6]{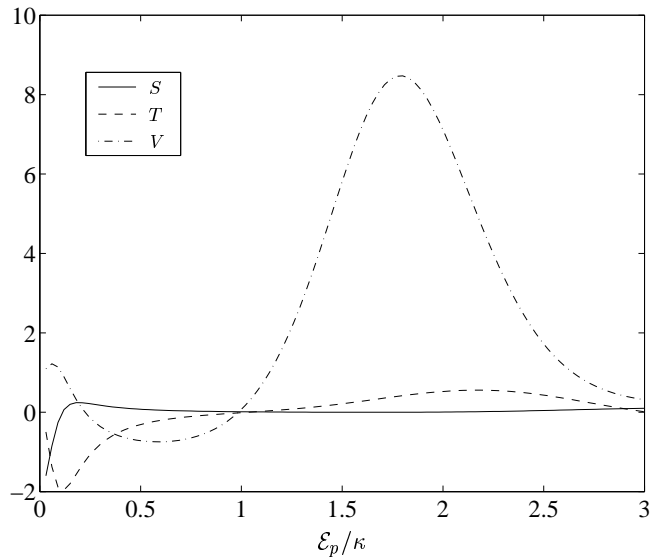}
  \end{center}
  \caption{Contributions to the correlation function $g^{(2)}(0)$ from the incoherent component of the intracavity field. Parameters are the same as in Fig.~\ref{fig:1}.}
  \label{fig:incoh}
\end{figure}

The contributions from the incoherent component of the intracavity field are shown in Fig.~\ref{fig:incoh}. The squeezing and intensity correlation parts are shown separately. The antibunching for weak fields comes from the squeezed fluctuations, which reduce the contribution from the coherent scattering. As the driving increases, the squeezing decreases, but the variance in intensity fluctuations becomes negative, so the remaining antibunching comes from the sub-Poissonian intensity fluctuations in the incoherent component of the field. For ${\mathcal E}_p \sim \kappa$, the squeezing contribution effectively vanishes, while $T$ and $V$ become positive, and antibunching disappears. Strong bunching clearly originates in the super-Poissonian intensity correlations, and the correlation function is dominated by $V$.

\section{Nonclassical Behaviour of the Correlation Function}
\label{sec:cauchy}

The presence of nonclassical effects in a driven atom-cavity system is a topic that has received much attention for many years. The system usually studied has been the canonical system of quantum optics -- a single two-level atom confined in an optical cavity. Photon statistics in the bad-cavity limit was thoroughly studied by Rice and Carmichael~\cite{Rice88}, who analyzed the sub-Poissonian statistics and photon antibunching in the cavity transmission, for the case of weak driving. Their analysis was extended by Carmichael {\it et al.}~\cite{Carmichael91} to a system containing $N$ two-level atoms. This analysis was further refined by Brecha {\it et al.}~\cite{Brecha99}. Clemens and Rice~\cite{Clemens00} have extended the consideration involving a single atom to include arbitrary driving field strength and dephasing. In their analysis, Clemens and Rice pay special attention to nonclassical effects known as `undershoots' and `overshoots'. These are related to the violation of inequalities that hold for classical correlations, in particular violations that occur not in the value of $g^{(2)}(\tau=0)$, but for certain time delays $\tau > 0$. The explanation for the undershoots has been given by Carmichael {\it et al.}~\cite{Carmichael91} in terms of quantum interference of probability amplitudes and collapse of the wavefunction.

The requirements for the classicality of the field correlations can be derived from the Cauchy-Schwartz inequality (see~\cite{Brecha99} and references therein), and expressed in terms of the second-order correlation function as
\begin{equation}
  \label{eq:cauchy}
  |g^{(2)}(\tau)-1| \leq |g^{(2)}(0)-1| \, .
\end{equation}
Values in excess of those allowed classically are called overshoots, while values below are called undershoots. Overshoots have been observed recently by Mielke {\it et al.}~\cite{Mielke98}.

In this context, it is of interest to see if the overshoots and/or undershoots can be found in the single-atom EIT-Kerr system under consideration. Photon antibunching, as an example of nonclassical photon statistics, has already been predicted~\cite{Rebic99}, and the effective two-level behaviour analyzed~\cite{Rebic99,Werner99,Rebic01}. In the present article, we have shown how the effects of self-homodyning of squeezed dipole radiation yields photon antibunching in the low to moderate driving limit. We have also shown that quantum interference between the probability amplitudes contributes to both strong antibunching and strong bunching, for weak and strong driving fields, respectively. Given this range of behaviours, we might therefore expect undershoots and overshoots to also occur in the single-atom EIT-Kerr system under suitable conditions.

\begin{figure}
  \begin{center}
    \includegraphics[scale=0.7]{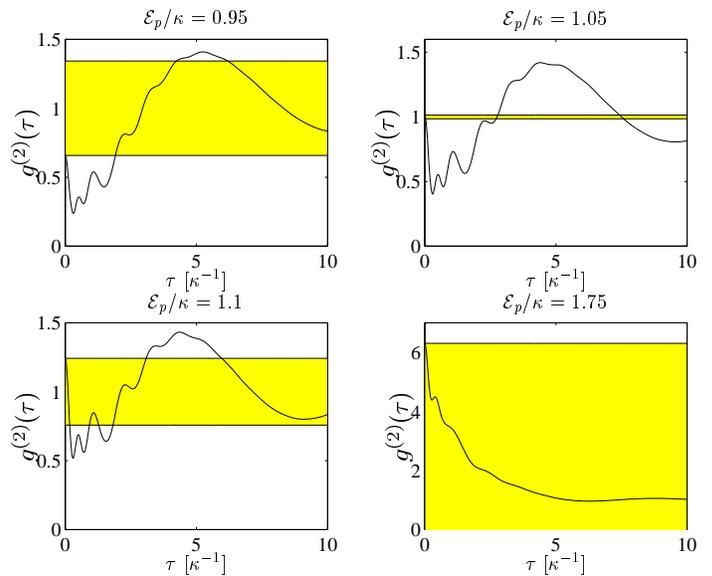}
  \end{center}
  \caption{Second order correlation functions vs. time delay for a different values of driving field strength. Shaded areas denote classically allowed values, calculated from Eq.~(\ref{eq:cauchy}).}
  \label{fig:corfuncs}
\end{figure}

Fig.~\ref{fig:corfuncs} shows correlation functions for several values of driving field strength. The values of driving have been chosen where nonclassical behaviour is expected to be found. For weak driving, where the antibunching is strong, the delay-time dependence of the correlation function is well-understood in terms of the effective two-level system. The interesting region is for those values of driving for which $g^{(2)}(0)$ increases through 1, the value for a coherent field. This is also the region in which the dynamics is well-described in terms of quantum interference and increased incidence of two-photon emissions.

Not surprisingly, this is also the range of parameters where largest violations of the classical inequalities occur. Since the increase in the value of $g^{(2)}(0)$ is due to the purely quantum effect of interference between probability amplitudes, non-classical behaviour of the correlations can be expected. The explanation of these effects given by Rice and Carmichael~\cite{Rice88} and Carmichael {\it et al.}~\cite{Carmichael91}, although in a different context, still holds. As shown in Sec.~\ref{sec:incoh}, self-homodyning of squeezed dipole radiation with the driving field occurs in the EIT-Kerr system in a similar manner to that for a two-level atom. An alternative explanation in terms of quantum interference of the driving field with the atomic polarization after the collapse of the wave function upon a photon detection event offers even more insight. This is best understood in the context of quantum trajectory theory.

\begin{figure}[!t]
  \begin{center}
    \includegraphics[scale=0.8]{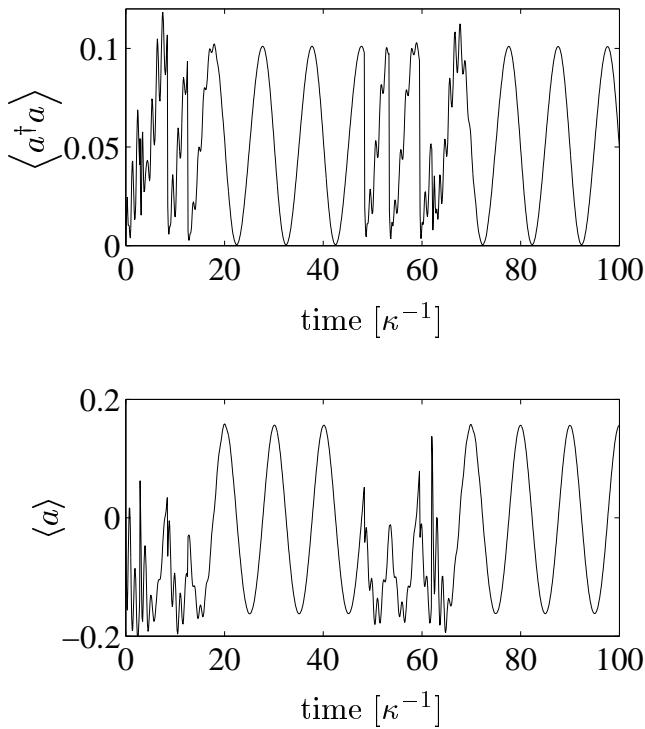}
  \end{center}
  \caption{Intracavity photon number and field amplitude in a typical realization of a single quantum trajectory, for ${\mathcal E}_p/\kappa = 1$.}
  \label{fig:traject}
\end{figure}

\begin{figure}[!t]
  \begin{center}
    \includegraphics[scale=0.6]{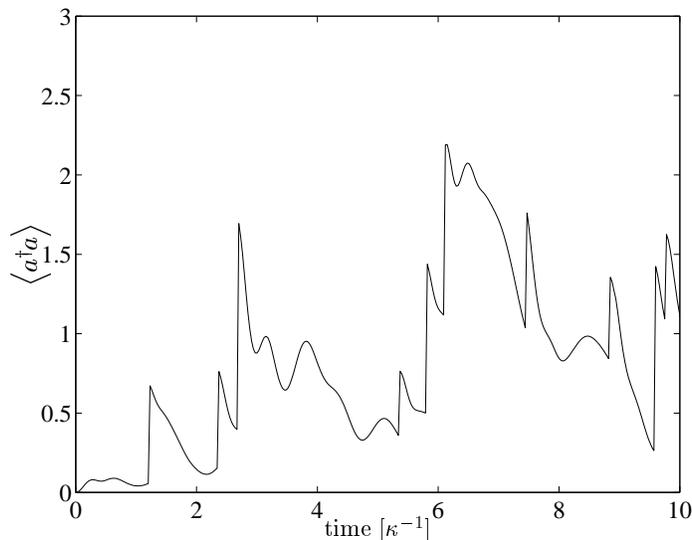}
  \end{center}
  \caption{Intracavity photon number in a typical realization of a single quantum trajectory, for ${\mathcal E}_p/\kappa = 1.75$.}
  \label{fig:addtraject}
\end{figure}

This explanation is illustrated in Fig.~\ref{fig:traject}, where single trajectory realizations for the intracavity field and photon number are plotted. We see that the collapses tend to occur in succession before the system returns to a (quasi-) steady state (which for this region of parameters happens after time $\gamma_j^{-1}$). Once it returns into a steady state, a few Rabi cycles pass before the next set of collapses occur. The value of $g^{(2)}(0)$ is determined by the ratio of the number of jumps upwards to number of jumps downwards in photon number, where jump upwards suggests that the detection of a photon increases the probability of detecting a second photon immediately afterwards. Naturally, at the value of driving where $g^{(2)}(0)$ peaks (see Fig.~\ref{fig:1}), collapses are almost exclusively upwards, as illustrated in Figure~\ref{fig:addtraject}. Undershoots appear as the consequence of a change in sign that the amplitude undergoes at the collapse~\cite{Carmichael91}. As the system returns towards its steady state the polarization becomes close in magnitude and opposite in sign to the driving field, producing a near-zero mean intracavity field, leading to the reduced detection probability for a second photon.

The overshoots can also be explained in terms of the collapses of the wave function. The detection of the first photon, emitted from the steady state situation, collapses the wave function of the system. The subsequent time evolution as the system returns to the steady state determines the photon correlations. For weak driving field, the probability for the second collapse to occur before the system returns to the steady state is extremely small, since it is proportional to the mean intracavity photon number. For stronger driving fields the probability for subsequent collapses increases, specifically due to the large correlations between first and third manifold states, as shown in Fig.~\ref{fig:popcoh}. Therefore, the probability for a second photon detection after some time $\tau<\gamma^{-1}$ increases as well, causing the correlation function overshoot. In experiment, such event pairs are the source of delayed coincidence counts. A third collapse is also likely to occur before the steady state is reached. However, the overshoot disappears (or significantly decreases) for all time delays $\tau$ after the first peak. This is expected, since the exact form of the wave function after the second collapse depends on the delay time between the second and third photon. Averaging over all possible realizations washes out the nonclassical effects due to different possible evolutions following the second collapse.

A stronger driving field causes more subsequent collapses to occur, and nonclassical correlations are completely washed out. We find that overshoots and undershoots vanish at driving strength ${\mathcal E}_p \approx 1.14 \kappa$.

\section{Conclusion}
\label{sec:conclusion}

We have presented an analysis of the properties of the photon statistics of light emitted by a single atom intracavity EIT-Kerr system. It was found that the statistics change abruptly as the driving field strength increases. Specifically, strong photon antibunching, dominant in the weak to moderate driving regimes, is replaced by a strong photon bunching in the output field for the stronger driving. We have identified the effect of quantum interference between the dressed states to be responsible for this sudden change, and presented an effective model explaining qualitative features of this behaviour.

Furthermore, we have analyzed contributions from the incoherent scattering to the system dynamics and found that the strong photon antibunching can be explained in terms of the self-homodyning of the incoherent intracavity component with the coherent component for weak driving, and in terms of reduced intensity fluctuations for moderate driving strengths. Strong bunching is the signature of super-Poissonian intensity fluctuations.

Finally, nonclassical behaviour of the correlation function was found, and the effects of undershoots and overshoots analyzed.

\begin{acknowledgments}
The authors would like to thank M. J. Collett and M. K. Olsen for many valuable comments and suggestions. This work was supported by the Marsden Fund of the Royal Society of New Zealand and The University of Auckland Research Committee.
\end{acknowledgments}

\appendix

\section{Stark Splitting of the Dressed States}
\label{sec:stark}

In this Appendix, we briefly review the effect of dynamic Stark splitting of dressed states, fully elaborated upon in Ref.~\cite{Rebic01}. 

Dressed states of the Hamiltonian~(\ref{eq:h0}) contain, among others, two states on exact cavity resonance, separated by energy $\hbar\omega_{cav}$. These are the ground state and a state belonging to the first excitation manifold. In the schematic depiction of Fig.~\ref{fig:eff6}, these are denoted by $|e_0\rangle$ and $|e_1\rangle$. Driving by the external field ${\mathcal E}_p$ enters through the effective Rabi frequency $\Omega_{01} = {\mathcal E}_p/\sqrt{1+g_1^2/\Omega_c^2}$. Resonant coupling of the two levels causes splitting of the two states into a symmetric and antisymmetric linear combination of the two
\begin{equation}
  |\psi_\pm\rangle = \left( |e_0\rangle \pm |e_1\rangle \right)/\sqrt{2} \, ,
\end{equation}
with corresponding energies
\begin{equation}
  \epsilon_\pm = \pm\sqrt{\Omega_{01}^2 - \left( \Gamma_0/2\right)^2} \, ,
\end{equation}
where $\Gamma_0 = \kappa/\left( 1+g_1^2/\Omega_c^2 \right)$ is the decay rate of the excited state $|e_1\rangle$. The splitting of energy levels occurs at the threshold values of driving field
\begin{equation}
  {\mathcal E}_p = \frac{\kappa/2}{\sqrt{1+g_1^2/\Omega_c^2}} \, .
\end{equation}
The two states $|\psi_\pm\rangle$ are the two states closest to the cavity resonance for the range of driving strengths considered in this article.

\section{Quantum Jumps and Density Matrix Elements}
\label{sec:jumprho}

In this Appendix, we derive jump terms for the effective master equation~(\ref{eq:master}), and the equations of motion for density matrix elements.

\subsection{Jump Terms}
\label{sec:jump}

We follow the notation of Carmichael~\cite{Carmichael93} and rewrite the effective master equation~(\ref{eq:master}) in a Lindblad form as $\dot{\rho}_{eff} = {\mathcal L}_{eff} \rho_{eff}$, where the Liouvillian superoperator can be divided into a part describing the free evolution between the jumps, ${\mathcal L}_{eff} - {\mathcal S}_{eff}$, and a part describing the jumps, ${\mathcal S}_{eff}$. Jump term ${\mathcal S}_{eff}\rho_{eff}$ arises from the equivalent jump term in the full master equation,
\begin{eqnarray}
  {\mathcal S}\rho &=& 2\kappa\, a\rho a^\dagger + 2\gamma_1\, \sigma_{12}\rho \sigma_{21} \nonumber \\
  &\ &+ 2\gamma_2\, \sigma_{32}\rho \sigma_{23} + 2\gamma_3\, \sigma_{34}\rho \sigma_{43} \label{eq:srhofull} \, .
\end{eqnarray}
Using the methods of Ref.~\cite{Rebic01}, and consistent with the truncation of the dressed states space (as in Fig.~\ref{fig:eff6}), we can write field and atomic collapse operators in terms of the bare states as
\begin{subequations}
  \label{eq:corresp}
\begin{eqnarray}
  a &=& |0,\, 1\rangle\langle 1,\, 1| + |0,\, 3\rangle\langle 1,\, 3| + |0,\, 4\rangle\langle 1,\, 4| \nonumber \\
    &\ & + \sqrt{2} \left( |1,\, 1\rangle\langle 2,\, 1| + |1,\, 2\rangle\langle 2,\, 2| + |1,\, 3\rangle\langle 2,\, 3|   \right) \nonumber \\
    &\ &+ \sqrt{3} |2,\, 1\rangle\langle 3,\, 1| \, , \\
  \sigma_{12} &=& |1,\, 1\rangle\langle 1,\, 2| + |2,\, 1\rangle\langle 2,\, 2| \, , \\
  \sigma_{32} &=& |1,\, 3\rangle\langle 1,\, 2| + |2,\, 3\rangle\langle 2,\, 2| \, , \\
  \sigma_{34} &=& |0,\, 3\rangle\langle 0,\, 4| + |1,\, 3\rangle\langle 1,\, 4| \, .
\end{eqnarray}
\end{subequations}
It can be deduced from the bare states that the collapse operators can be expressed as linear combinations of the following polariton operators: $a \propto \left\{ p_1,\, p_2,\, p_3,\, p_{24},\, p_{34},\, p_{25},\, p_{35} \right\}$; $\sigma_{12},\, \sigma_{34} \propto \left\{ p_2,\, p_3,\, p_{24},\, p_{34},\, p_{25},\, p_{35} \right\}$ and $\sigma_{32} \propto \left\{ p_2^\dagger p_2,\, p_3^\dagger p_3,\, p_4^\dagger p_4,\, p_5^\dagger p_5,\, p_2^\dagger p_3,\, p_3^\dagger p_2,\, p_{j4}^\dagger p_{j5},\, p_{j5}^\dagger p_{j4} \right\}$. Note that the polariton jump terms arising from the atomic jumps associated with the operator $\sigma_{32}$ (i.e. proportional to $\gamma_2$) couple dressed states within the same manifold. This is possible since the associated atomic transition is not coupled to the cavity mode, but to the classical field $\Omega_c$, and therefore jumps occurring in this atomic decay channel do not switch between the adjacent manifolds. 

Using the correspondences above, one can write all of the jump terms in the polariton basis, thus making the transition ${\mathcal S}\rho \rightarrow {\mathcal S}_{eff}\rho_{eff}$. This transformation can be viewed as a rotation of a truncated basis of a superoperator. The coefficients $\Gamma_{ijkl}$ of the jump terms are given in Table~\ref{tab:coeff}.

\subsection{Equations of Motion}
\label{sec:rhoij}

The equations of motion for the density matrix elements in the basis schematically shown in Fig.~\ref{fig:eff6} can be found from the master equation~(\ref{eq:master}). There is a total of 35 equations, since population conservation can be used to eliminate one of them. The equations for populations are
\begin{subequations}
  \label{eq:populations}
\begin{eqnarray}
  \dot{\rho}_{00} &=& 2\Gamma_0 \rho_{11} + \Omega_0 \left( \rho_{01} + \rho_{10} \right) \, , \\
  \dot{\rho}_{11} &=& -2\Gamma_0 \rho_{11} + \Gamma_{1122} \rho_{22} + \Gamma_{1133} \rho_{33} \nonumber \\
  &\ &- \Omega_0 \left( \rho_{01} + \rho_{10} \right) \nonumber \\
  &\ &+ \Omega_{12}\rho_{12}+\Omega_{12}^*\rho_{21} + \Omega_{13}\rho_{13}+\Omega_{13}^*\rho_{31} \, , \\
  \dot{\rho}_{22} &=& -2\left( \Gamma_{22} +\Gamma_{2222} \right)\rho_{22}  \nonumber \\
  &\ &-\Gamma_{23} \left( \rho_{32} + \rho_{23} \right) + 2\Gamma_{45} \left( \rho_{45} + \rho_{54} \right) \nonumber \\
  &\ &-  \Omega_{12}\rho_{12}-\Omega_{12}^*\rho_{21} + \Gamma_{2233} \rho_{33} + \Gamma_{2244} \rho_{44} + \Gamma_{2255} \rho_{55} \nonumber \\
  &\ &+  \Omega_{24} \left( \rho_{24} + \rho_{42} \right) +  \Omega_{25} \left( \rho_{25} + \rho_{52} \right) \, , \\
  \dot{\rho}_{33} &=& -2\left(\Gamma_{33} + \Gamma_{3333} \right) \rho_{33}  \nonumber \\
  &\ &-\Gamma_{23} \left( \rho_{32} + \rho_{23} \right) + 2\Gamma_{45} \left( \rho_{45} + \rho_{54} \right) \nonumber \\
  &\ &-  \Omega_{13}\rho_{13}-\Omega_{13}^*\rho_{31} + \Gamma_{3322} \rho_{22} + \Gamma_{3344} \rho_{44} + \Gamma_{3355} \rho_{55} \nonumber \\
  &\ &+ \Omega_{34} \left( \rho_{34} + \rho_{43} \right) +  \Omega_{35} \left( \rho_{35} + \rho_{53} \right) \, , \\
  \dot{\rho}_{44} &=& -2\left( \Gamma_{44} + \Gamma_{4444} \right)\rho_{44} + \Gamma_{4455} \rho_{55} - 2\Gamma_{45} \left( \rho_{45} + \rho_{54} \right) \nonumber \\
  &\ &-\Omega_{24} \left( \rho_{24} + \rho_{42} \right) - \Omega_{34} \left( \rho_{34} + \rho_{43} \right) \, , \\
  \dot{\rho}_{55} &=& -2\left( \Gamma_{55} +\Gamma_{5555} \right)\rho_{55} + \Gamma_{5544} \rho_{44} - 2\Gamma_{45} \left( \rho_{45} + \rho_{54} \right) \nonumber \\
  &\ &-\Omega_{25} \left( \rho_{25} + \rho_{52} \right) - \Omega_{35} \left( \rho_{35} + \rho_{53} \right) \, .
\end{eqnarray}
\end{subequations}
The equations for coherences are
\begin{subequations}
  \label{eq:coherences}
\begin{eqnarray}
  \dot{\rho}_{01} &=& -\Gamma_0 \rho_{01} + 2\Gamma_{02}\rho_{12} + 2\Gamma_{03}\rho_{13} \nonumber \\
  &\ & + \Omega_0 \left( \rho_{11} - \rho_{00} \right) + \Omega_{12}\rho_{02} + \Omega_{13}\rho_{03} \\
  \dot{\rho}_{02} &=& -\left(\Gamma_{22}+\Gamma_{2222}-i\epsilon_2 \right) \rho_{02} -\Gamma_{23}\rho_{03} \nonumber \\
  &\ &+ \Gamma_{0214}\rho_{14} + \Gamma_{0215}\rho_{15} \nonumber \\
  &\ & + \Omega_0 \rho_{12} - \Omega_{12}^*\rho_{01} + \Omega_{24}\rho_{04} + \Omega_{25}\rho_{05} 
\end{eqnarray}
\begin{eqnarray}
  \dot{\rho}_{03} &=& -\left(\Gamma_{33}+\Gamma_{3333}-i\epsilon_3 \right) \rho_{03} -\Gamma_{23}\rho_{02} \nonumber \\
  &\ &+ \Gamma_{0314}\rho_{14} + \Gamma_{0315}\rho_{15} \nonumber \\
  &\ & + \Omega_0 \rho_{13} - \Omega_{13}^*\rho_{01} + \Omega_{34}\rho_{04} + \Omega_{35}\rho_{05} \\
  \dot{\rho}_{04} &=& -\left(\Gamma_{44}+\Gamma_{4444}+i\epsilon_4 \right) \rho_{04} -\Gamma_{45}\rho_{05} \nonumber \\
  &\ & - \Omega_{24}\rho_{02} - \Omega_{34}\rho_{03} +\Omega_0\rho_{14} \\
  \dot{\rho}_{05} &=& -\left(\Gamma_{55}+\Gamma_{5555}+i\epsilon_5 \right) \rho_{05} -\Gamma_{45}\rho_{04} \nonumber \\
  &\ & - \Omega_{25}\rho_{02} - \Omega_{35}\rho_{03} +\Omega_0\rho_{12} \\
  \dot{\rho}_{12} &=& -\left(\Gamma_0+\Gamma_{22}+\Gamma_{2222}-i\epsilon_2 \right) \rho_{12} - \Gamma_{23}\rho_{13} \nonumber \\
  &\ & + \Gamma_{1224}\rho_{24} + \Gamma_{1225}\rho_{25} + \Gamma_{1234}\rho_{34} + \Gamma_{1235}\rho_{35} \nonumber \\
  &\ & + \Omega_{12}^* \left( \rho_{22} - \rho_{11} \right) \nonumber \\
  &\ &- \Omega_0 \rho_{02} +\Omega_{13}^* \rho_{32} + \Omega_{24}\rho_{14} + \Omega_{25}\rho_{15} \\
  \dot{\rho}_{13} &=& -\left(\Gamma_0+\Gamma_{33}+\Gamma_{3333}-i\epsilon_3 \right) \rho_{13} - \Gamma_{23}\rho_{12} \nonumber \\
  &\ & + \Gamma_{1324}\rho_{24} + \Gamma_{1325}\rho_{25} + \Gamma_{1334}\rho_{34} + \Gamma_{1335}\rho_{35} \nonumber \\
  &\ & + \Omega_{13}^* \left( \rho_{33} - \rho_{11} \right) \nonumber \\
  &\ &- \Omega_0 \rho_{03} +\Omega_{12}^* \rho_{23} + \Omega_{34}\rho_{14} + \Omega_{35}\rho_{15} \\
  \dot{\rho}_{14} &=& -\left(\Gamma_0+\Gamma_{44}+\Gamma_{4444}-i\epsilon_4 \right) \rho_{14} - \Gamma_{45}\rho_{15} - \Omega_0 \rho_{04} \nonumber \\
  &\ &+ \Omega_{24}\rho_{12} + \Omega_{34}\rho_{13}  +\Omega_{12}^* \rho_{24} +\Omega_{13}^* \rho_{34}\\
  \dot{\rho}_{15} &=& -\left(\Gamma_0+\Gamma_{55}+\Gamma_{5555}-i\epsilon_5 \right) \rho_{15} - \Gamma_{45}\rho_{14} - \Omega_0 \rho_{05} \nonumber \\
  &\ &+ \Omega_{25}\rho_{12} + \Omega_{35}\rho_{13}  +\Omega_{12}^* \rho_{25} +\Omega_{13}^* \rho_{35}\\
  \dot{\rho}_{23} &=& -\left[\Gamma_{22}+\Gamma_{2222}+\Gamma_{33}+\Gamma_{3333}+\Gamma_{2233}-i\left(\epsilon_2+\epsilon_3\right) \right] \rho_{23} \nonumber \\
  &\ &  -\Gamma_{23}\left(\rho_{22}+\rho_{33}\right) + \Gamma_{2344}\rho_{44} + \Gamma_{2355}\rho_{55} + \Gamma_{2332}\rho_{32} \nonumber \\
  &\ & + \Gamma_{2345}\rho_{45} + \Gamma_{2354}\rho_{54}+\Omega_{12}^* \rho_{13} +\Omega_{13}^* \rho_{21} \nonumber \\
  &\ & + \Omega_{34} \rho_{24} + \Omega_{35} \rho_{25} + \Omega_{24}\rho_{43} + \Omega_{25}\rho_{53} \\
  \dot{\rho}_{24} &=& -\left[\Gamma_{22}+\Gamma_{2222}+\Gamma_{44}+\Gamma_{4444}+\Gamma_{2244}-i\left(\epsilon_2+\epsilon_4\right) \right] \rho_{24}  \nonumber \\
  &\ & -\Gamma_{23}\rho_{34}-\Gamma_{45}\rho_{25} + \Gamma_{2435}\rho_{35}  \nonumber \\
  &\ & - \Omega_{24}\left( \rho_{44} - \rho_{22} \right)-\Omega_{12}\rho_{14} - \Omega_{34}\rho_{23}+\Omega_{25}\rho_{54}\\
  \dot{\rho}_{25} &=& -\left[\Gamma_{22}+\Gamma_{2222}+\Gamma_{55}+\Gamma_{5555}+\Gamma_{2255}-i\left(\epsilon_2+\epsilon_5\right) \right] \rho_{25}  \nonumber \\
  &\ & -\Gamma_{23}\rho_{35}-\Gamma_{45}\rho_{24} + \Gamma_{2534}\rho_{34}  \nonumber \\
  &\ & - \Omega_{25}\left( \rho_{55} - \rho_{22} \right)-\Omega_{12}\rho_{15} - \Omega_{35}\rho_{23}+\Omega_{24}\rho_{45} \\
  \dot{\rho}_{34} &=& -\left[\Gamma_{33}+\Gamma_{3333}+\Gamma_{44}+\Gamma_{4444}+\Gamma_{3344}-i\left(\epsilon_3+\epsilon_4\right) \right] \rho_{34}  \nonumber \\
  &\ & -\Gamma_{23}\rho_{24}-\Gamma_{45}\rho_{35}+ \Gamma_{3425}\rho_{25}  \nonumber \\
  &\ & - \Omega_{34}\left( \rho_{44} - \rho_{33} \right)-\Omega_{13}\rho_{14} + \Omega_{35}\rho_{54}-\Omega_{24}\rho_{32} \\
  \dot{\rho}_{35} &=& -\left[\Gamma_{33}+\Gamma_{3333}+\Gamma_{55}+\Gamma_{5555}+\Gamma_{3355}-i\left(\epsilon_3+\epsilon_5 \right) \right] \rho_{35}  \nonumber \\
  &\ & -\Gamma_{23}\rho_{25}-\Gamma_{45}\rho_{34}+ \Gamma_{3524}\rho_{24}  \nonumber \\
  &\ & - \Omega_{35}\left( \rho_{55} - \rho_{33} \right)-\Omega_{13}\rho_{15} + \Omega_{34}\rho_{45}-\Omega_{25}\rho_{32} \\
  \dot{\rho}_{45} &=& -\left[\Gamma_{44}+\Gamma_{4444}+\Gamma_{55}+\Gamma_{5555}+\Gamma_{4455}-i\left(\epsilon_4+\epsilon_5\right) \right] \rho_{45} \nonumber \\
  &\ & - \Gamma_{45}\left( \rho_{44} + \rho_{55} \right) + \Gamma_{4554}\rho_{55} \nonumber \\
  &\ & -\Omega_{24}\rho_{25} - \Omega_{34}\rho_{35}-\Omega_{25}\rho_{42} -\Omega_{35}\rho_{43} \, .
\end{eqnarray}
\end{subequations}
Damping coefficients $\Gamma_0$ and $\Gamma_{ij}$ have been evaluated in general in Ref.~\cite{Rebic01}. Coefficients $\Gamma_{ijkl}$ are related to the jump operators and are given in Table~\ref{tab:coeff}. The source of the different terms in these equations should be apparent from the earlier discussion. For example, damping emerging from jump terms is denoted as $\Gamma_{ijkl}$, while damping coming from the effective Hamiltonian~(\ref{eq:heff6}) is denoted as $\Gamma_0$ and $\Gamma_{ij}$.

It is straightforward to prove that the following equalities hold:
\begin{subequations}
  \label{eq:conds}
\begin{eqnarray}
  2\left(\Gamma_{22} + \Gamma_{2222}\right) &=& \Gamma_{1122} + \Gamma_{3322} \, , \\
  2\left(\Gamma_{33} + \Gamma_{3333}\right) &=& \Gamma_{1133} + \Gamma_{2233} \, , \\
  2\left(\Gamma_{44} + \Gamma_{4444}\right) &=& \Gamma_{2244} + \Gamma_{3344} + \Gamma_{5544} \, , \\
  2\left(\Gamma_{55} + \Gamma_{5555}\right) &=& \Gamma_{2255} + \Gamma_{3355} + \Gamma_{4455} \, ,
\end{eqnarray}
\end{subequations}
preserving the population conservation condition $\sum_j \rho_{jj} = 1$, i.e. $\sum_j \dot{\rho}_{jj} = 0$.
  
\begin{table*}[!t]
\begin{center}
  \fbox{
\begin{tabular} {|| c | c || c | c ||} \hline \hline \\
  $ijkl$ & $\Gamma_{ijkl}/2$ & $ijkl$ & $\Gamma_{ijkl}/2$ \\ \hline \hline
  2233 & $\gamma_2 |\mu_2^{(2)}|^2 |\beta_3^{(2)}|^2$ & 2244 & $\kappa |w_{24}|^2 + \gamma_1 |\mu_2^{(3)}|^2 |\alpha_2^{(2)}|^2 + \gamma_3 |\mu_2^{(2)}|^2 |\nu_2^{(3)}|^2$ \\
  \hline
  3322 & $\gamma_2 |\mu_3^{(2)}|^2 |\beta_2^{(2)}|^2$ & 3344 & $\kappa |w_{34}|^2 + \gamma_1 |\mu_2^{(3)}|^2 |\alpha_3^{(2)}|^2 + \gamma_3 |\mu_3^{(2)}|^2 |\nu_2^{(3)}|^2$ \\
  \hline
  4455 & $\gamma_2 |\mu_2^{(3)}|^2 |\beta_3^{(3)}|^2$ & 1334 & $\kappa w_{13}w_{34}^* + \gamma_1 \alpha_1^{(1)*} \mu_2^{(3)*} \alpha_3^{(2)} \beta_3^{(2)} + \gamma_3 \mu_3^{(1)*} \nu_2^{(3)*} \mu_3^{(2)} \nu_3^{(2)}$ \\
  \hline
  0215 & $\kappa w_0 w_{25}^*$ & 1224 & $\kappa w_{12}w_{24}^* + \gamma_1 \alpha_1^{(1)*} \beta_2^{(2)} \alpha_2^{(2)} \mu_2^{(3)*} + \gamma_3 \mu_3^{(1)*} \mu_2^{(2)} \nu_2^{(3)*} \nu_2^{(2)}$ \\
  \hline
  0314 & $\kappa w_0 w_{34}^*$ & 1225 & $\kappa w_{12}w_{25}^* + \gamma_1 \alpha_1^{(1)*} \beta_2^{(2)} \alpha_2^{(2)} \mu_3^{(3)*} + \gamma_3 \mu_3^{(1)*} \mu_2^{(2)} \nu_3^{(3)*} \nu_2^{(2)}$ \\
  \hline
  0315 & $\kappa w_0 w_{35}^*$ & 1324 & $\kappa w_{12}w_{34}^* + \gamma_1 \alpha_1^{(1)*} \beta_2^{(2)} \alpha_3^{(2)} \mu_2^{(3)*} + \gamma_3 \mu_3^{(1)*} \mu_3^{(2)} \nu_2^{(3)*} \nu_2^{(2)}$ \\
  \hline
  5544 & $\gamma_2 |\mu_3^{(3)}|^2 |\beta_2^{(3)}|^2$ & 1335 & $\kappa w_{13}w_{35}^* + \gamma_1 \alpha_1^{(1)*} \beta_3^{(2)} \alpha_3^{(2)} \mu_3^{(3)*} + \gamma_3 \mu_3^{(1)*} \mu_3^{(2)} \nu_3^{(3)*} \nu_3^{(2)}$  \\
  \hline
  0214 & $\kappa w_0 w_{24}^*$ & 2332 & $\gamma_2 \beta_2^{(2)*}\beta_3^{(2)}\mu_2^{(2)*}\mu_3^{(2)}$ \\
  \hline
  2534 & $\gamma_2  \mu_2^{(2)*} \beta_2^{(3)*} \beta_3^{(2)} \mu_3^{(3)}$ & 2435 & $\gamma_2  \mu_2^{(2)*} \beta_3^{(3)*} \beta_3^{(2)} \mu_2^{(3)}$ \\
  \hline
  3425 & $\gamma_2  \mu_3^{(2)*} \beta_3^{(3)*} \beta_2^{(2)} \mu_2^{(3)}$ & 3524 & $\gamma_2  \mu_3^{(2)*} \beta_2^{(3)*} \beta_3^{(3)} \mu_2^{(3)}$ \\
  \hline
   2255 & $\kappa |w_{25}|^2 + \gamma_1 |\mu_2^{(3)}|^3 |\alpha_2^{(2)}|^2 + \gamma_3 |\mu_2^{(2)}|^2 |\nu_3^{(3)}|^2$ & 3355 & $\kappa |w_{35}|^2 + \gamma_1 |\mu_3^{(3)}|^2 |\alpha_3^{(2)}|^2 + \gamma_3 |\mu_3^{(2)}|^2 |\nu_3^{(3)}|^2$ \\
   \hline
   2355 & $\kappa w_{25}w_{35}^* + \gamma_1 \alpha_2^{(2)*} |\mu_3^{(3)}|^2 \alpha_3^{(2)} + \gamma_3 \mu_2^{(2)*} |\nu_3^{(3)}|^2 \mu_3^{(2)}$ & 1235 & $\kappa w_{13}w_{25}^* + \gamma_1 \alpha_1^{(1)*} \beta_3^{(2)} \alpha_3^{(2)} \mu_3^{(3)*} + \gamma_3 \mu_3^{(1)*} \mu_3^{(2)} \nu_3^{(3)*} \nu_3^{(2)}$ \\
   \hline
   1234 & $\kappa w_{13}w_{24}^* + \gamma_1 \alpha_1^{(1)*} \beta_3^{(2)} \alpha_2^{(2)} \mu_2^{(3)*} + \gamma_3 \mu_3^{(1)*} \mu_2^{(2)} \nu_2^{(3)*} \nu_3^{(2)}$ & 1325 & $\kappa w_{12}w_{35}^* + \gamma_1 \alpha_1^{(1)*} \beta_2^{(2)} \alpha_3^{(2)} \mu_3^{(3)*} + \gamma_3 \mu_3^{(1)*} \mu_3^{(2)} \nu_3^{(3)*} \nu_2^{(2)}$ \\
   \hline
   2344 & $\kappa w_{24}w_{34}^* + \gamma_1 \alpha_2^{(2)*} \mu_2^{(3)} \alpha_3^{(2)} \mu_2^{(3)*} + \gamma_3 \mu_2^{(2)*} \mu_3^{(2)} |\nu_2^{(3)}|^2$ & 2345 & $\kappa w_{24}w_{35}^* + \gamma_1 \alpha_2^{(2)*} \mu_2^{(3)} \alpha_3^{(2)} \mu_3^{(3)*} + \gamma_3 \mu_2^{(2)*} \mu_3^{(2)} \nu_2^{(3)} \nu_3^{(3)*}$ \\
   \hline
   2354 & $\kappa w_{25}w_{34}^* + \gamma_1 \alpha_2^{(2)*} \mu_2^{(3)*} \alpha_3^{(2)} \mu_3^{(3)} + \gamma_3 \mu_2^{(2)*} \mu_3^{(2)} \nu_2^{(3)*} \nu_3^{(3)}$ & 4554 & $\gamma_2  \mu_2^{(3)*} \beta_2^{(3)*} \beta_3^{(3)} \mu_3^{(3)}$ \\
   \hline
   2222 & $(\gamma_2/2) |\mu_2^{(2)}|^2 |\beta_2^{(2)}|^2$ & 3333 & $(\gamma_2/2) |\mu_3^{(2)}|^2 |\beta_3^{(2)}|^2$ \\
   \hline
   4444 & $(\gamma_2/2) |\mu_2^{(3)}|^2 |\beta_2^{(3)}|^2$ & 5555 & $(\gamma_2/2) |\mu_3^{(3)}|^2 |\beta_3^{(3)}|^2$ \\
   \hline
   2233 & $(\gamma_2/2) |\mu_2^{(2)}|^2 |\beta_3^{(2)}|^2$ & 2244 & $\kappa |w_{24}|^2 + \gamma_1 |\alpha_2^{(2)}|^2 |\mu_2^{(3)}|^2 + \gamma_3 |\mu_2^{(2)}|^2 |\nu_2^{(3)}|^2$ \\
   \hline
   2255 & $\kappa |w_{25}|^2 + \gamma_1 |\alpha_2^{(2)}|^2 |\mu_3^{(3)}|^2 + \gamma_3 |\mu_2^{(2)}|^2 |\nu_3^{(3)}|^2$ & 3344 & $\kappa |w_{34}|^2 + \gamma_1 |\alpha_3^{(2)}|^2 |\mu_2^{(3)}|^2 + \gamma_3 |\mu_3^{(2)}|^2 |\nu_2^{(3)}|^2$ \\
   \hline
   3355 & $\kappa |w_{35}|^2 + \gamma_1 |\alpha_3^{(2)}|^2 |\mu_3^{(3)}|^2 + \gamma_3 |\mu_3^{(2)}|^2 |\nu_3^{(3)}|^2$ & 4455 & $(\gamma_2/2) |\mu_2^{(3)}|^2 |\beta_3^{(3)}|^2$ \\
   \hline \hline
\end{tabular}
}
\end{center}
\caption{Expressions for the `jump' coefficients. Here, $w_{ij} = \Omega_{ij}/{\mathcal E}_p$, and the parameters $\alpha,\, \beta,\, \mu$ and $\nu$ are given in~\cite{Rebic01}.}
\label{tab:coeff}
\end{table*}

\end{document}